\documentclass[letter]{aa}
\usepackage{natbib,twoopt}
\usepackage{comment}
\usepackage{hyperref} %% to avoid \citeads line fills
\hypersetup{
    colorlinks=true,
    linkcolor=red,
    filecolor=magenta,      
    urlcolor=cyan,
    pdftitle={Overleaf Example},
    pdfpagemode=FullScreen,
    citecolor=blue,
    breaklinks=true
    }
\bibpunct{(}{)}{;}{a}{}{,}             %% natbib format for A&A and ApJ
\makeatletter
  \newcommandtwoopt{\citeads}[3][][]{\href{http://adsabs.harvard.edu/abs/#3}%
    {\def\hyper@linkstart##1##2{}%
     \let\hyper@linkend\@empty\citealp[#1][#2]{#3}}}
  \newcommandtwoopt{\citepads}[3][][]{\href{http://adsabs.harvard.edu/abs/#3}%
    {\def\hyper@linkstart##1##2{}%
     \let\hyper@linkend\@empty\citep[#1][#2]{#3}}}
  \newcommandtwoopt{\citetads}[3][][]{\href{http://adsabs.harvard.edu/abs/#3}%
    {\def\hyper@linkstart##1##2{}%
     \let\hyper@linkend\@empty\citet[#1][#2]{#3}}}
  \newcommandtwoopt{\citeyearads}[3][][]%
    {\href{http://adsabs.harvard.edu/abs/#3}
    {\def\hyper@linkstart##1##2{}%
     \let\hyper@linkend\@empty\citeyear[#1][#2]{#3}}}
\makeatother
\usepackage[toc,page]{appendix}
\usepackage{adjustbox}
\usepackage{xcolor}
\usepackage{graphicx}
\usepackage{txfonts}
\usepackage{tabularx, blindtext}
\usepackage{rotating}
\usepackage{pdflscape}
\usepackage{inputenc}
\usepackage{caption}
\usepackage{longtable}
\usepackage{lscape}
\DeclareCaptionFormat{continued}{#1 (continued)#2#3\par}
\DeclareGraphicsExtensions{.jpg,.png, .pdf}
\usepackage[utf8]{inputenc}
\usepackage{titlesec}

\begin{document} 

%\title{Where are the super metal-rich Bulge globular clusters?}
%\subtitle{Revealing an inconsistency}
\title{VVVX survey dusts off a new intermediate-age star cluster in the Milky Way disk}
%\subtitle{bjkhkjhk}
   \author{E. R. Garro \inst{1} 
          \and
          D. Minniti\inst{2,3,4}
               \and
          J. Alonso-García \inst{5,6}
             \and
          J.~G.~Fernández-Trincado\inst{7}
             \and
          M. Gómez\inst{2}
             \and
          T. Palma\inst{8}
              \and 
          R. K. Saito\inst{3}
            \and
          C. Obasi\inst{7}
          }
   \institute{
   ESO - European Southern Observatory, Alonso de Cordova 3107, Vitacura, Santiago, Chile
   \and 
   Instituto de Astrofísica, Depto. de Ciencias Físicas, Facultad de Ciencias Exactas, Universidad Andres Bello, Av. Fernandez Concha 700, Las Condes, Santiago, Chile
   \and
 Vatican Observatory, Vatican City State, V-00120, Italy
 \and
 Departamento de F\'isica, Universidade Federal de Santa Catarina, Trindade 88040-900, Florian\'opolis, Brazil
 \and
 Centro de Astronomía (CITEVA), Universidad de Antofagasta, Av. Angamos 601, Antofagasta, Chile
\and
Millennium Institute of Astrophysics (MAS), Nuncio Monseñor Sotero Sanz 100, Of. 104, Providencia, Santiago, Chile
 \and
 Instituto de Astronom\'ia, Universidad Cat\'olica del Norte, Av. Angamos 0610, Antofagasta, Chile
 \and 
 Observatorio Astronómico, Universidad Nacional de Córdoba, Laprida 854, X5000BGR Córdoba, Argentina
}
  \date{Received: XX; Accepted: YY}
 
\abstract
   {%CONTEXT. 
In the last decade, many new star clusters have been discovered in heavily obscured regions of the Milky Way bulge and disk.}
   {%AIMS.
Our primary long-term objective is to seek out additional star clusters in the poorly studied regions of the Milky Way, where detections pose significant challenges. The aim of this pursuit is to finalize the Milky Way's globular and open cluster system census and to gain a comprehensive understanding of both the formation and evolution of these systems and our Galaxy as a whole.}
  % methods heading (mandatory)
   {%METHODS.
We report the discovery of a new star cluster, named Garro~03. We investigated this new target using a combination of near-infrared and optical databases. We employed the VISTA Variables in the Via Láctea Survey and Two Micron All Sky Survey data in the  near-infrared, and the \textit{Gaia} Data Release 3 and the DECam Plane Survey datasets in the optical passband. We constructed density maps and vector proper motion diagrams in order to highlight our target. We performed a photometrical analysis in order to derive its main physical parameters.}
  % results heading (mandatory)
   {%RESULTS.
 Garro~03 is located at equatorial coordinates RA~=~14:01:29.3 and Dec~=~$-$65:30:57.0. From our photometric analysis we find that this cluster is not heavily affected by extinction with $A_{Ks}=0.25 \pm 0.04$ mag and $A_G = 1.54\pm 0.02$ mag. It is located at heliocentric distance of $14.1\pm0.5$ kpc, which places Garro~03 at $10.6$ kpc from the Galactic centre and Z = $-0.89$ kpc below the Galactic plane. We also calculated the mean cluster proper motion of ($\mu_{\alpha}^{\ast},\mu_{\delta}) = (-4.57\pm 0.29,\ -1.36\pm 0.27$) mas yr$^{-1}$. We derived an age of 3 Gyr and metallicity [Fe/H]~$= -0.5\pm 0.2$ by the isochrone-fitting method, employing the PARSEC models. The total luminosity was derived in the  $K_s$ and V bands, finding $M_{Ks} = -6.32\pm 1.10$ mag and $M_V =-4.06$ mag. Finally, the core and tidal radii were measured constructing the Garro~03 radial density profile and fitting the King model. We obtained $r_c = 3.07\pm 0.98$ pc and $r_t = 19.36\pm 15.96$ pc, respectively.}
  {% CONCLUSION.
We photometrically confirm the cluster nature for Garro~03, located in the Galactic disk. It is a distant, low-luminosity, metal-rich star cluster of intermediate age. We also searched for possible signatures (streams or bridges) between Garro~03 and Garro~01, but we exclude a companionship with the present analysis. We need spectroscopic data to classify it as an old open cluster or a young globular cluster, and to understand {its origin}.}
\keywords{Galaxy: disk – Galaxy: stellar content – Galaxies: star clusters: general – Infrared: stars – Surveys}
   \maketitle
    
%\onecolumn
\section{Introduction}
\label{sec:Introduction}
The search for new star clusters in the Milky Way (MW) is revealing ever more  suitable candidates, especially in those regions heavily affected by differential reddening and crowding, such as the Galactic plane and bulge. In this context, 37 recently analysed globular clusters (GCs) have been listed in the compilation made by \citet[][and references therein]{2024arXiv240503068B,2024arXiv240505055G}, or numerous open cluster (OC) candidates (see e.g. \citealt[][among others)]{2022A&A...661A.118C,2022A&A...660A...4H}, making a significant step forward in our understanding of the MW structure and evolution. This effort holds tremendous potential for several reasons. The first  is the completion of the Galactic GC \citep{Minniti_2017} and OC \citep{2024arXiv240305143H} census. For instance, with thorough spectroscopic follow-ups,  new GCs could contribute to a potential increase of  20\% (or more)  in the total number of known MW GCs in the coming years. The second is   exploring the dissolution and destruction of a GC and OC \citep{Gnedin1997, 2008ApJ...685..247B, 2014MNRAS.441..150B}. Given their faint nature, these clusters likely lost stars over their lifetimes. However, the extent of this loss and the processes involved, including the formation of tidal tails, remain unclear. Additionally, comprehending the processes of mass loss can provide insights into the potential links between stars lost by clusters and field stars, shedding light on the nature and extent of this connection \citep{Kruijssen2012,Adamo2015}. The third reason is  for investigating the origin and formation of GCs. By distinguishing between in situ or ex situ formation scenarios \citep{2018A&A...620A.194R, 2019A&A...630L...4M, 10.1093/mnras/stad3920, 10.1093/mnras/stad3046}, we can uncover whether these clusters were accreted or formed during the early stages of MW formation. Additionally, they shed light on processes, such as wet mergers, that  contribute gas to our Galaxy and that potentially lead to the formation of a second generation of cluster population \citep{2019MNRAS.489.3269C, 10.1093/mnras/stab3682}. Fourth, we need to  understand how to distinguish old OCs from young GCs since there is no clear separation between these two classes. Finally, as shown in \cite{2024arXiv240505055G}, increasing the number of MW GCs can help us  to understand the formation and evolution of the MW itself, upgrading the luminosity function, age, and metallicity distributions, and other tools used to study our Galaxy. \\

For these reasons, one of the primary objectives of the VISTA Variables in the Via Láctea (VVV) Team revolves around the exploration of new star clusters within the VVV Survey \citep{2010NewA...15..433M, 2012A&A...537A.107S} and its eXtension (VVVX - \citealt{2024arXiv240616646S}) based on observations at the VISTA 4m telescope. However, the search for new objects is very difficult; in particular, in the crowded disk and bulge regions, which are obscured by dust, numerous clusters may have lost mass over time or initially formed with fewer stars. This is due to the cluster initial mass function \citep{2010A&A...512A..79H, 2015A&A...582A..93S,10.1093/mnras/stad631}, which suggests the formation of many low-mass clusters, far more than  we observe today among old GCs. Detecting certain star clusters can indeed be challenging due to their low luminosity. \\%Additionally, the presence of dust and field stars likely obscures these objects, further complicating their detection.

In this letter we present the discovery of a new star cluster candidate in the Galactic disk, as described in Sect. \ref{sec:discovery}. In Sect. \ref{sec:Methodoly and results} we explain the decontamination procedure performed to obtain a clean catalogue with the most probable cluster stars, and we derive the main physical parameters %(reddening and extinction, mean cluster proper motions PMs, size, heliocentric and Galactocentric distances, age and metallicity, integrated luminosity, and structural parameters) 
of our target for the first time. In Sect. \ref{sec:discussions} we explore the potential companionship of our discovery with another candidate, named Garro~01, given their strikingly similar characteristics. We conclude our findings in Sect. \ref{sec:conclusions}.

\section{Discovery and observational datasets}
\label{sec:discovery}
\begin{figure*}[htpb]
\centering
\includegraphics[width=6cm, height=6cm]{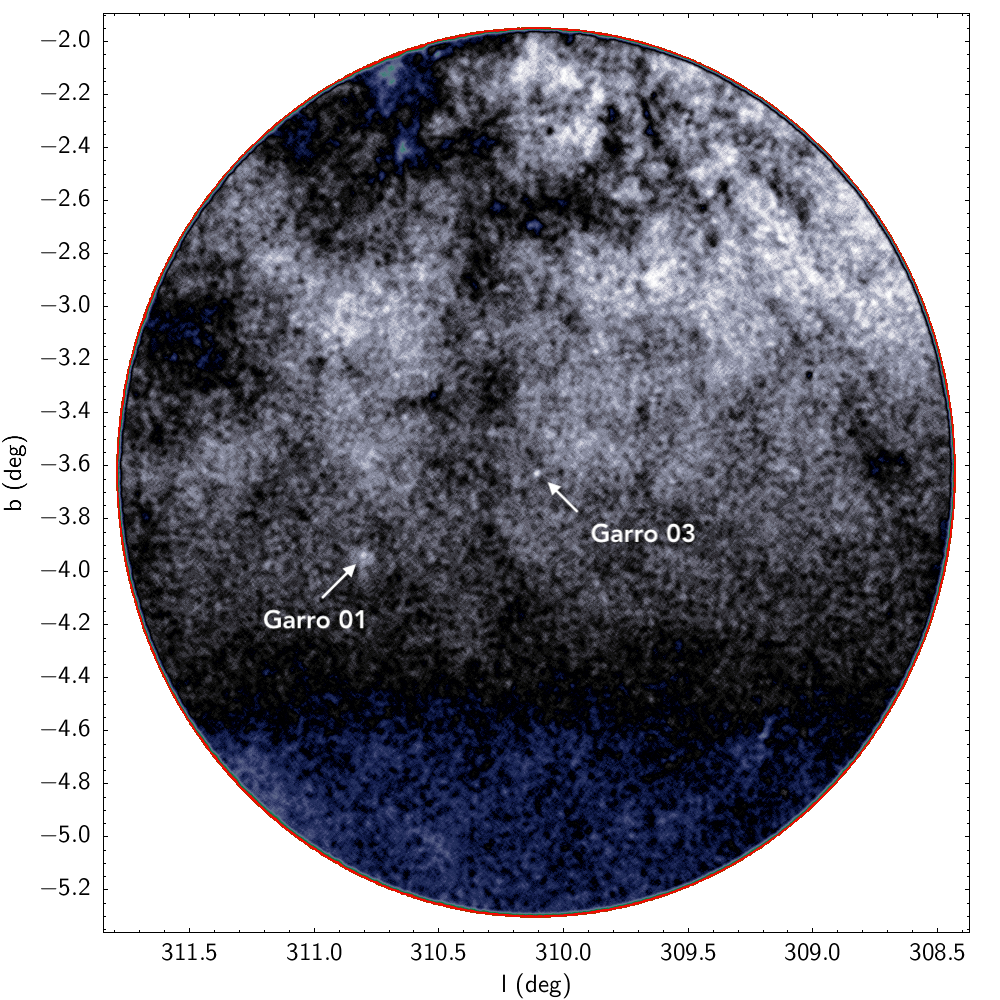} 
\includegraphics[width=6cm, height=6cm]{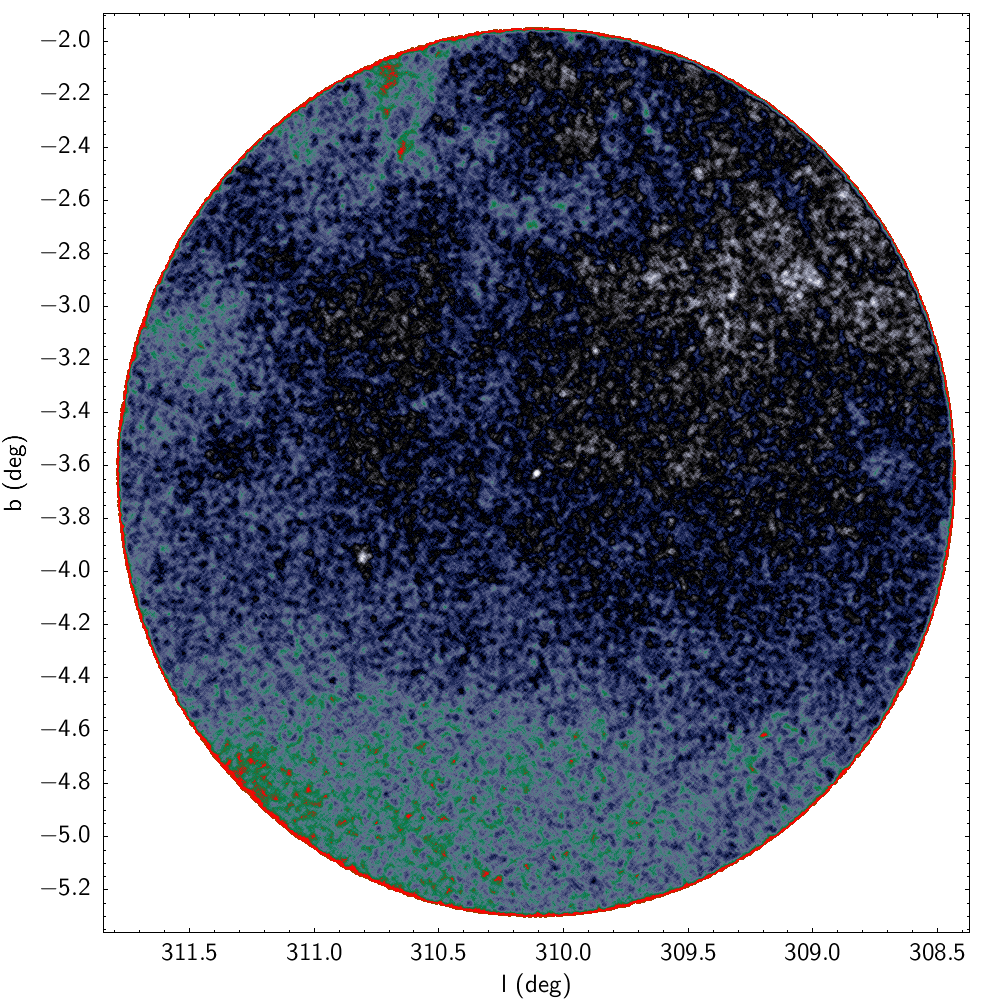} 
\includegraphics[width=6cm, height=6cm]{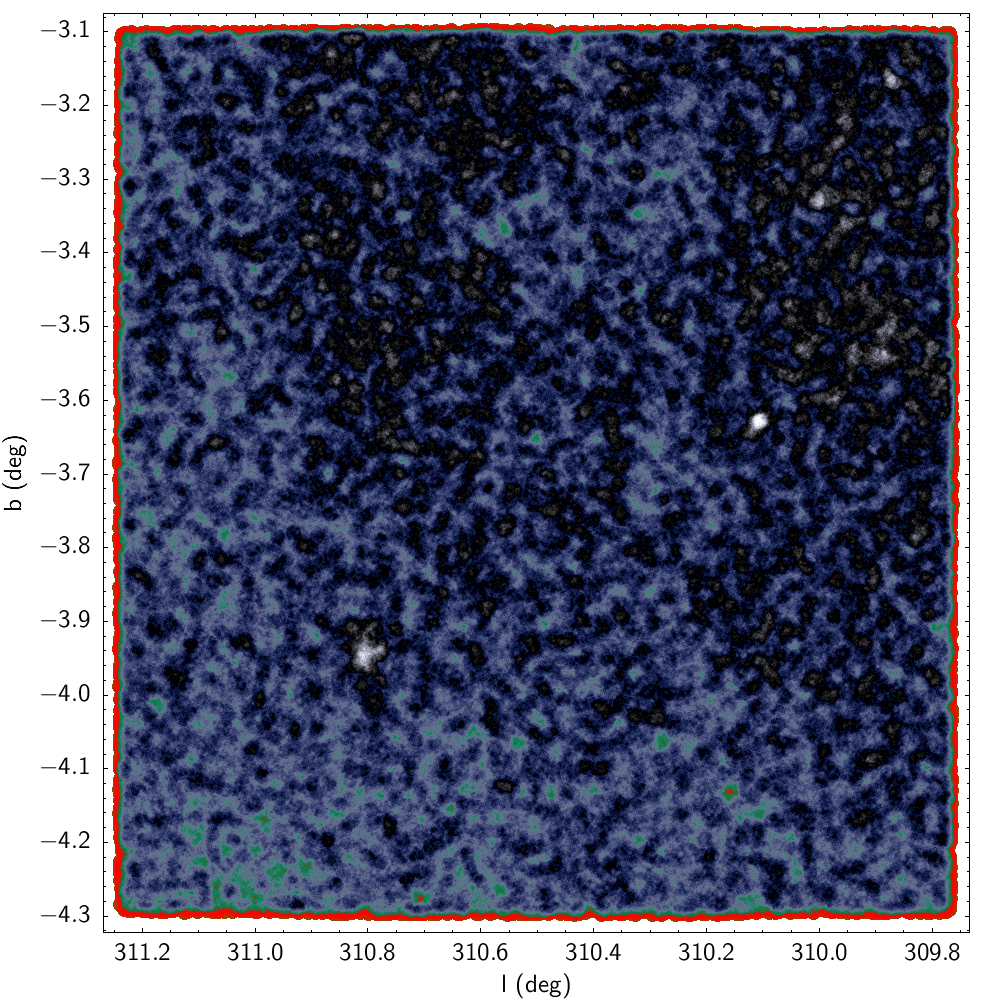} 
\caption{\textit{Gaia} DR3 stellar density map for a $r=100'$ field in the Galactic coordinates, centred on Garro~03 target.  On the left panel, we draw two arrows to indicate the location of two overdensities, named Garro~01 and Garro~03. In this case, we did not make any cuts. In the middle panel, we make the cuts: \texttt{plx} < 0.5, \texttt{ruwe} < 1.4, -6.0 < \texttt{pmra} < -2.0 mas yr$^{-1}$, -2.0 < \texttt{pmdec} < 0 mas yr$^{-1}$, in order to clean the density map and to better highlight the two overdensities. The right panel shows a zoomed-in view, and we used a combination of \textit{Gaia} DR3+VVVX data. For all panels, the density's colours increase from the black (lower density) to white (higher density).}
\label{fig:twoGCs}
\end{figure*}

While doing a thorough inspection of the low-latitude field surrounding the Circinus galaxy (D = 4 Mpc) searching for associated background GCs and dwarf galaxies (\citealt{2024MNRAS.529.3075O,2023A&A...670A..18O}, respectively), we found a few foreground Galactic star cluster candidates. For example, we confirmed a new star cluster candidate in the Galactic disk, designated \object{Garro~01} (RA = 14:09:00.0 and Dec = $-$65:37:12 -- \citealt{Garro_2020}). It was indicated as a GC by \cite{Garro_2020} with an age~$\approx 11$ Gyr and [Fe/H]~$=-0.7$, but it was classified as an old OC by \cite{2023MNRAS.526.1075P} with an age $\approx 4$ Gyr and [Fe/H] $=-0.3$. Thus, its classification   remains ambiguous. \\
An examination of the VVVX tile $e0656$, which houses this cluster, revealed a distinct and separated overdensity of red giant stars at equatorial coordinates RA~=~14:01:29.3 %(equivalent to 210.372 deg)
and Dec~=~$-$65:30:57.0, 
%(equivalent to $-$65.516 deg), 
corresponding to l~$=310.111$ deg and b~$=-3.625$ deg, suggesting the presence of another star cluster. Through visual inspection (Fig. \ref{fig:twoGCs}), we took precautions to ensure that the observed overdensity was not merely a transient feature sensitive to parameter adjustments. To achieve this, we refined the catalogue by applying the following criteria: \texttt{plx} $< 0.5$ (to clean it from nearby stars), renormalized unit error ruwe \texttt{ruwe} $< 1.4$ (high-quality astrometric parameter), $-6.0 <$ \texttt{pmra} $< -2.0$ mas yr$^{-1}$, $-2.0 <$ \texttt{pmdec} $< 0$ mas yr$^{-1}$ (as derived in Sect. \ref{ssec:parameters} the two clusters have similar PMs). This cleaning process accentuated the clarity of both overdensities. Notably, the central overdensity  exhibits a distinctive circular shape, consistent with the characteristics expected for a star cluster. Consequently, we report the discovery of a star cluster candidate, hereafter named \object{Garro~03}, situated at $47'$ far from Garro~01, and shown in Fig. \ref{fig:twoGCs}. \\

As done in our previous works \citep{Garro_2020, Garro2021b, Garro2021a, Garro2022_Garro02, Garro2023}, we investigated this new candidate in order to first confirm the existence of the cluster, and then to determine its nature. To do this, we combined both near-infrared (NIR) and optical datasets, adopting a $0.5''$ matching radius. We used the NIR dataset from the VVVX and Two Micron All-Sky Survey (2MASS, \citealt{2006AJ....131.1163S}). The VVVX is a public ESO survey, acquired with the VISTA Infrared CAMera (VIRCAM) at the  4.1m wide-field Visible and Infrared Survey Telescope for Astronomy (VISTA, \citealt{2010Msngr.139....2E}). The point spread function (PSF) photometry was extracted as described by \cite{2018A&A...619A...4A}. Astrometry was calibrated to the \textit{Gaia} Data Release 3 (DR3; \citealt{2018A&A...616A...1G}) astrometric reference system, while the  photometry was calibrated to the VISTA magnitude system \citep{Gonzalez_Fernandez2018} against the 2MASS using a globally optimized model of frame-by-frame zero points plus an illumination correction. \\ 
We used NIR 2MASS data to reach magnitudes brighter than K$_s=11$, which are saturated in the VVVX images. We treated the VVVX and 2MASS data separately, but we scaled them to the same photometric system \citep{Gonzalez_Fernandez2018}, and applied correction offsets of $\Delta K_s< 0.009$ mag and $\Delta J < 0.050$ mag.\\
{We also checked the \textit{Gaia} DR3 images. For instance, we provide density maps (Fig. \ref{fig:twoGCs}), applying different cuts, using simply the \texttt{TOPCAT} software and its `density' mode, we found} a clear excess of stars at the same coordinates. As done for Garro~01 \citep{Garro_2020}, we analysed this target using this optical dataset, benefitting from the  accurate astrometry and PMs, selecting stars with \texttt{plx} > 0.5 mas (equivalent to $D<3$ kpc - \citealt{BailerJones2018}) and \texttt{ruwe} < 1.4 \citep{Fabricius2021,GaiaCollaboration2020}. As explained in Sect. \ref{ssec:decontamination and membership}, we used \textit{Gaia} PMs to separate cluster stars from field stars. Finally, we integrated the optical DECam Plane Survey (DECaPS, \citealt{2018ApJS..234...39S, 2023ApJS..264...28S}) in order to reach the upper part of the main-sequence turn-off (MSTO).

\section{Methodology and results}
\label{sec:Methodoly and results}
\begin{figure*}[htpb]
\centering
\includegraphics[width=15cm, height=10cm]{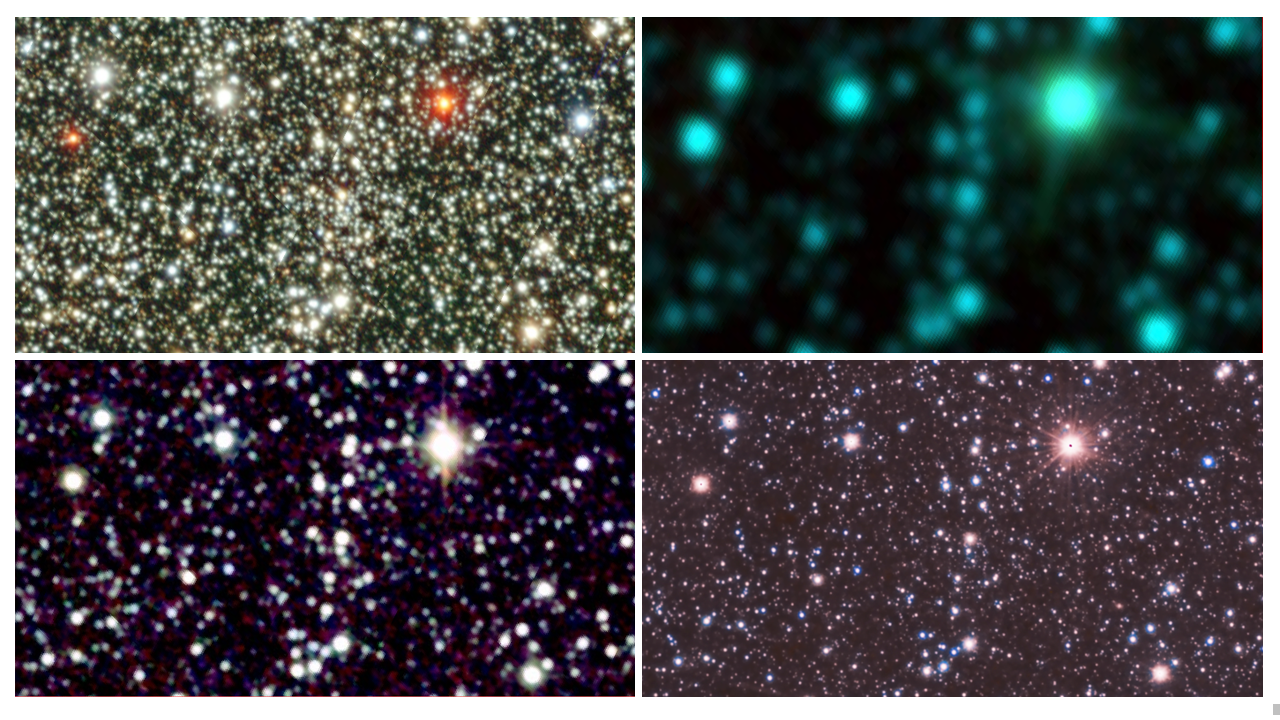} 
\caption{DECaPS (top left), ALL WISE (top right), 2MASS (on bottom left), and VVVX (bottom right) images centred on Garro~03. The field of view is $5.5'\times 3.0'$, oriented along the Equatorial coordinates, with the longitude increasing to the left and the latitude increasing to the top.}
\label{fig:images}
\end{figure*}

We wanted to ascertain whether the identified overdensity holds the characteristics of a genuine star cluster or if it might be a mere grouping of stars or statistical fluctuations in the projected stellar density within the plane of the sky, which were leading us astray. Therefore, adhering to the methodology outlined by \cite{Garro2022_Garro02}, we commenced our investigation by examining images captured at various wavelengths (see Fig. \ref{fig:images}). Notably, the overdensity is situated within a densely populated celestial region, prompting us to employ the Gaussian kernel density estimate (KDE). This statistical approach allows us to discern overdensities from lower-densities more objectively than relying solely on visual inspections. The application of the KDE method revealed a centralized and circular overdensity, as evident in Fig. \ref{fig:KDEposition} {(see Appendix \ref{appendix:A} for a detailed explanation)}. Furthermore, the iso-density contours illustrate a gradual decrease in density with increasing distance from the centre, characteristic of stellar clusters. Instead, these features are not visible for a field star population selected $4'$  from the cluster. These tests provide compelling evidence supporting the identification of a new star cluster candidate.

Although a spectroscopic analysis remains imperative to strengthen our discovery, the subsequent sections (Sects. \ref{ssec:decontamination and membership} and \ref{ssec:parameters}) will delve into the characterization of Garro~03. Our aim is to categorize this object as a promising candidate for either a GC or an OC, although the differences between these two classes have not yet been clarified, especially when comparing old OCs and young GCs.

\subsection{Decontamination procedure and cluster membership}
\label{ssec:decontamination and membership}
\begin{figure*}[htpb]
\centering
\includegraphics[width=5cm, height=5cm]{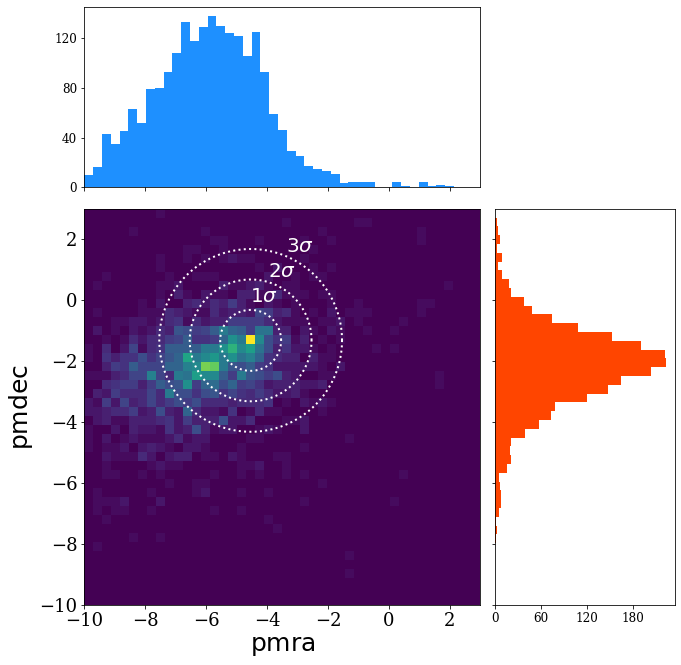} 
\includegraphics[width=6cm, height=5cm]{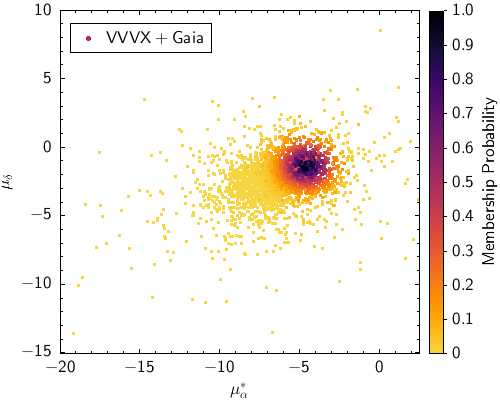} 
\includegraphics[width=6cm, height=5cm]{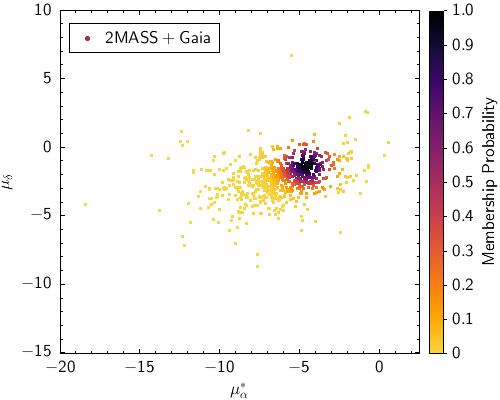} 
\caption{VPM diagrams. Left panel: VPM diagram for Garro~03 cluster as a 2D histogram, displaying both the PMRA (in blue) and PMDEC (in red) histograms separately. The yellow and light green colours depict overdensities, whereas the dark green and blue colours represent lower densities. We indicate with dotted white circles the 1$\sigma$, 2$\sigma$, and 3$\sigma$ centred on the mean cluster PMs. Middle and right panels: VPM diagrams for the \textit{Gaia}+VVVX and \textit{Gaia}+2MASS samples, respectively. The differences in colours reproduce the PM membership probability, as depicted in the colour-bar.}
\label{fig:PM}
\end{figure*}

\begin{figure*}[htpb]
\centering
\includegraphics[width=12cm, height=6cm]{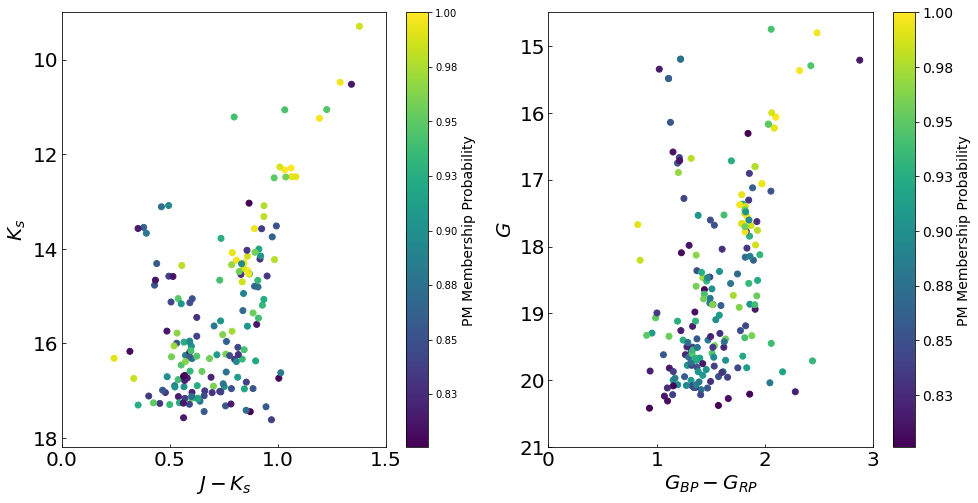} 
\caption{CMDs in optical (right panel) and NIR (left panel) passbands for the new cluster Garro~03. We show only selected stars with PM membership probability $>80\%$, as specified in the colour bar.}
\label{fig:cmd_probability}
\end{figure*}

As anticipated above, we followed the same procedure described in \cite{Garro2022_Garro02}.  Here we present a concise overview of the main steps undertaken.\\
Initially, we excluded nearby stars and focused on those exhibiting \texttt{ruwe}$<1.4$, ensuring the reliability of the astrometric solutions. Subsequently, employing KDE iso-density contours (Fig. \ref{fig:KDEposition}), we determined the size of  Garro~03  by selecting the region with the highest density around the centre within $r\sim 3'$. Lastly, we executed the PM decontamination procedure, employing two distinct methods. Firstly,  we constructed a vector PM diagram, represented as a 2D histogram in Fig. \ref{fig:PM}. A distinct peak in the distribution was evident, and through the application of the  sigma-clipping technique, we computed the mean cluster PM: $\mu_{\alpha}^{\ast} = -4.57 \pm 0.29$ mas yr$^{-1}$ and $\mu_{\delta}= -1.36 \pm 0.27$ mas yr$^{-1}$. Secondly, we determined PM probability membership for all stars, utilizing both the \textit{Gaia}+VVVX and \textit{Gaia}+2MASS samples. We favoured stars with PMs within $1$ mas yr$^{-1}$ of the mean cluster PM (Fig. \ref{fig:PM}). These two methods collectively facilitated the identification of stars with a likelihood of membership exceeding $80\%.$\footnote{We consider a very high PM membership probability because our main aim is to obtain a CMD that is as clean as possible in order to better identify the main evolutionary sequences, and to obtain robust physical parameters.} Subsequently, we constructed the colour-magnitude diagram (CMD) for Garro~03, incorporating all member stars (Fig. \ref{fig:cmd_probability}). Notably, a distinct red clump (RC) is discernible at $K_{s}= 14.33$ mag. Nevertheless, the CMD reveals an elongated blue sequence, indicating an over-representation of stars along the main sequence of the Galactic disk, which represent a very small percentage of contaminants. Our compilation lacks the fainter stars that constitute the turn-off point of the cluster's main sequence as they fall just below the detection limit of our magnitude measurements. Alternatively, the blue stars represent the cluster MSTO, and this may be an OC of intermediate age.

\subsection{Calculation of main physical parameters}
\label{ssec:parameters} 
The determination of the main physical parameters allows us to  characterize of our target. We  confirm its nature as a distant cluster for the first time. As already said, we applied the same procedure as explained in \cite{Garro2022_Garro02}.\\

Initially, we used the reddening maps provided by \cite{2011ApJ...737..103S} to determine optical extinction $A_V$ and reddening $E(B-V)$. The derived values were $A_V = 1.97 \pm 0.02$ mag and $E(B-V) = 0.64\pm 0.02$ mag. Utilizing the conversion relationships from \cite{2011ApJ...737..103S}, we determined the NIR extinction as $A_{Ks} = 0.22 \pm 0.02$ mag and the excess colour as $E(J-K_{s})=0.29 \pm 0.02$ mag, employing an extinction coefficient $R_{Ks}=0.75$ \citep{1989ApJ...345..245C}. Employing these parameters, we computed a distance modulus of $(m-M)_{0} = 15.72 \pm 0.05$ mag, equivalent to a distance of $D = 13.93 \pm 0.2$ kpc.

Another method for determining reddening and extinction involved applying the relations proposed by \cite{Babusiaux_2018}. This approach yielded a reddening $E(BP-RP)=0.77\pm 0.01$ mag and an extinction $A_G = 1.54 \pm 0.02$ mag. Consequently, we derived a distance modulus of $(m-M)_{0} = 15.75 \pm 0.02$ mag and a distance of $D = 14.11 \pm 0.2$ kpc. In the NIR, relations from \cite{2018A&A...609A.116R}, leveraging the RC position, were used. This resulted in an excess colour of $E(J-K_s)=0.24\pm 0.04$ mag and a reddening of $A_{Ks}=0.18\pm 0.03$ mag. Ultimately, a distance modulus of $(m-M)_{0}=15.75 \pm 0.04$ mag and a distance of $D = 14.16 \pm 0.4$ kpc were determined.

The consistent agreement among the various distance estimates positions Garro~03 at a heliocentric distance of 14.1 kpc. With $R_{\odot}=8.2$ kpc \citep{Gravity2019}, this corresponds to a  galactocentric distance of $R_G = 10.63$ kpc. Furthermore, the cluster is located at a distance below the Galactic plane of $Z=-0.89$ kpc, calculated using the relation $Z= D \times \sin(b)$, assuming $Z_{\odot}= 0$ kpc.

Upon calculating the mean cluster PMs and distances, we derived its tangential velocities as $V_{T}^{RA}\approx -305.68 $ km s$^{-1}$ and $V_{T}^{Dec}\approx -90.92$ km s$^{-1}$. These positional and kinematic parameters collectively suggest that Garro~03 is a disk cluster.\\

\begin{figure*}[htpb]
\centering
\includegraphics[width=10cm, height=14cm]{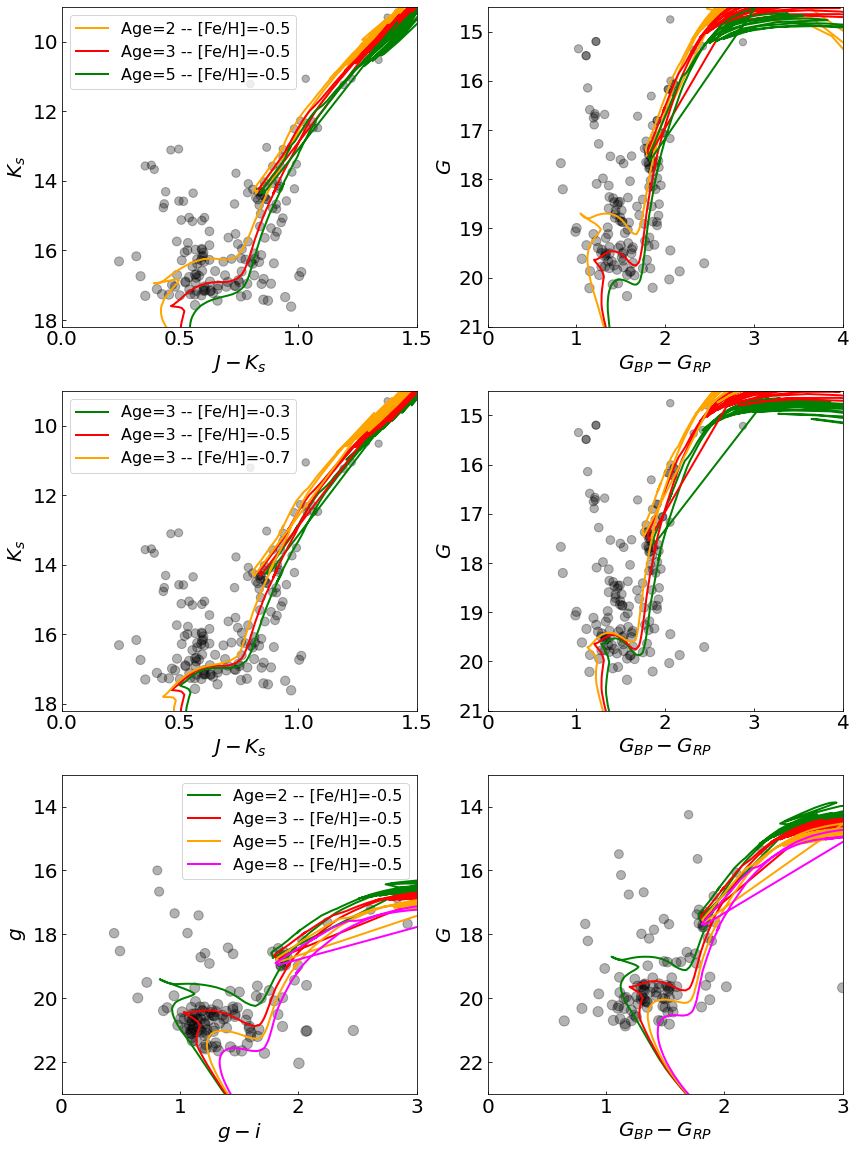} 
\caption{Optical and NIR CMDs, as shown in Fig. \ref{fig:cmd_probability}. We fit a family of PARSEC isochrones: on the top panels, we keep fixed the metallicity ([Fe/H] = $-0.5$) and we change the age (Age = $2,3, 5$ Gyr), contrary in the middle panels, we keep fixed the Age = 3 Gyr, and we change the metallicity ([Fe/H] = $-0.3,\ -0.5,\ -0.7$), as also specified by the legend. In the bottom panels, we show deeper DECaPS+\textit{Gaia} DR3 CMDs, where the top of the main sequence is reached at $g\sim 20.8$, used to derive a robust age estimate. In this case, we fixed the [Fe/H] $=-0.5$ and we change the age $=2, 3, 5, 8$ Gyr. }
\label{fig:cmd_isochrones}
\end{figure*}

Once the reddening, extinction, and distance modulus have been derived, we can employ them to estimate the metallicity and age using the isochrone fitting method. The main challenge in this case is the degeneracy between metallicity and age since the MSTO is not reached if we use the VVVX+\textit{Gaia} data, and thus the estimation of the age is complicated. Hence, we used the cross-correlated DECaPS and \textit{Gaia} DR3 data to construct a deep CMD, reaching the upper part of the MSTO (Fig. \ref{fig:cmd_isochrones}). The PARSEC best fit yields an age of 3-3.5 Gyr. Independently, we derived the metallicity estimates using the slope$_{RGB} -$ [Fe/H] linear relations proposed by \cite{Cohen_2015} in the NIR. We derived the RGB-slope as the line connecting two points along the RGB at the HB level and $\sim 2.5$ mag brighter. The resulting metallicity [Fe/H] $\approx -0.35$. These two approaches help to limit the age-metallicity degeneracy problem. Therefore, starting from these values, we fitted the PARSEC isochrones\footnote{\url{http://stev.oapd.inaf.it/cgi-bin/cmd}} downloading a family of isochrones in both the NIR and optical passbands. Figure \ref{fig:cmd_isochrones} shows the  best-fitting isochrones, finding [Fe/H] $=-0.5 \pm 0.2$ (in agreement with the Cohen estimate) and age $=3\pm 1$ Gyr, obtained from the overplotting of the isochrones that most closely fit simultaneously the points in the NIR and optical CMDs. \\

Another parameter that we derived is the integrated luminosity. We face the same problem that appeared in our previous studies \citep{Garro_2020, Garro2021a, Garro2022_Garro02} since the  main-sequence faint stars are missing in our compilation. However, as demonstrated in \cite{Garro2021b,Garro2021a,Garro2022_Garro02} and \cite{Kharchenko2016}, we can calculate a reasonable total luminosity since the first 10-12 brighter stars accumulate more than half of the integrated luminosity of a cluster; therefore, our estimate is a lower limit of the real cluster total luminosity. We can derive this parameter by comparing the luminosity of Garro~03 with those of known and well-characterized Galactic GCs, following the same method as in \cite{Garro2021a}. In this case, we first derived the absolute magnitude, measuring the cluster total flux. We obtained $M_{Ks} = -6.32 \pm 1.10$ mag, which is equivalent to $M_V = -3.82$ mag, if we assume the typical colour of $(V-K_s)=2.5$ mag (e.g. \citealt{Barmby2000, Conroy_2010}). Consequently, we made a comparison with NGC~6642, NGC~6626, NGC~6540, and NGC~6558 GCs since their metallicity is similar to that of    Garro~03. This allowed us to quantify the fraction of luminosity missed; we added this amount to the Garro~03 luminosity, obtaining a final value of $M_V=-4.06$ mag. \\

\begin{figure*}[htpb]
\centering
\includegraphics[width=7cm, height=5cm]{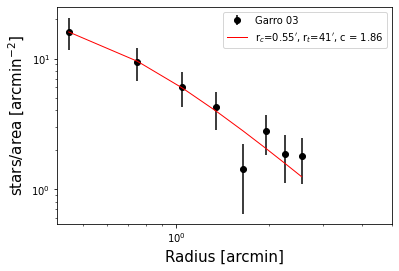} 
\includegraphics[width=7cm, height=5cm]{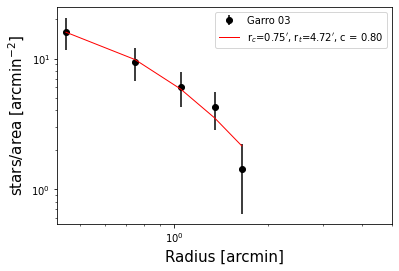} 
\caption{Radial density profile of Garro~03, considering eight circular annuli (on the left) and excluding the last three annuli, considered background level residuals (on the right). In both panels, we depict with the red line the best-fit of the King model, which provides the main structural parameters specified in the legend.}
\label{fig:RDP}
\end{figure*}

Finally, we computed the radial density profile for Garro~03. We followed the same procedure described by \cite{Garro2022_Garro02}, with the main aim of deriving the main structural parameters:  core radius, tidal radius, and concentration. Since the radial density profile may be affected by the determination of the centre of the cluster, we first checked that the centre coordinates of the system were accurately determined for a proper determination of the radial density profile. Given the very small difference (only a few arcseconds) among the  first  and  second  centre determinations, we opted for the first, original, centre value (listed in Table \ref{table}). Therefore, we divided our sample into eight circular annuli, increasing the radii by 0.3' out to an outer radius of 4.8'. Each bin contains between 6 and 14 stars, indicating that our statistics are low, due to the fact that we are considering only evolved stars with very high PM membership probability. We calculated the density per bin as the number of stars (N) divided by the area (A), while the error bars were computed as $e = \sqrt{N/A}$. We subtracted the background level, fixed at 0.5 stars/arcmin$^{2}$, although it may be slightly higher (Fig. \ref{fig:RDP}). We fitted the King model \citep{1962AJ.....67..471K} to provide the main structural parameters. Considering all eight circular annuli, the best-fit King model yields a core radius $r_c = 0.55'$ (equivalent to 2.25 pc at a distance of 14.1 kpc), a tidal radius $r_t=41'$ (equivalent to 168.16 pc at a distance of 14.1 kpc), and a concentration $c= \log_{10}(r_t / r_c) = 1.86$. We find that these values are not reasonable scientifically, and therefore we excluded the external three points (Fig. \ref{fig:RDP}), considering them as background level residuals, but including the lowest point in our calculation. In this case, the best-fit King model returns a $r_c = 0.75' \pm 0.24'$ (equivalent to $3.07\pm 0.98$ pc), a $r_t=4.72' \pm 3.89$ (equivalent to $19.36\pm 15.96$ pc), and a $c = 0.80$, which are consistent with the typical OC and GC radii listed in \cite{2022A&A...659A..59T} and in the  2010 edition of the  \cite{1996AJ....112.1487H} compilations, respectively. %We are affected by low statistics and the radial density profile is not well-resolved, but comparing our values with the typical structural parameters of known Galactic GCs, we can suggest that Garro~03 is a core-collapse cluster. 

\section{Discussions}
\label{sec:discussions}
%As detailed in \cite{2024arXiv240505055G} and references therein, numerous new star cluster (both OCs and GCs) candidates have emerged within the Galactic bulge and disk in recent years. This ongoing discovery underscores that the MW GC and OCs census remains a dynamic endeavour. The pursuit of these new clusters represents a significant milestone, with the potential to reshape our understanding of the history, formation, and evolution of the MW.\\
We explored the relationship between Garro~01 and Garro~03 due to their proximity and similar heliocentric and galactocentric distances. These clusters also share comparable traits, such as  low luminosity, compact sizes, and matching PMs, prompting us to question whether their histories are intertwined. %We acknowledge the necessity of spectroscopic follow-up to substantiate our conjecture, since through orbit reconstruction and studying their chemical evolution, we will be able to discern whether these clusters have encountered analogous dynamic destruction processes.

Currently, our investigation focuses on identifying signatures such as bridges or streams linking clusters. Although we employed diverse algorithms, primarily based on positions and PMs of all the stars in our catalogue, interpreting whether these clusters may be companions is an intricate task. To do this, we first calculated the PM membership probability (as done in Sect. \ref{ssec:decontamination and membership}) for all stars present in the tile $e0656$; then we selected only stars with PM membership probability > 50\%; and  finally we applied the KDE technique, as done in Fig. \ref{fig:KDEposition}. Figure \ref{fig:companion} again shows the two overdensities, representing the two clusters, but we do not detect any features that can suggest a connection between Garro~01 and Garro~03. This complexity may arise because the background field might overlap with the `imaginary' bridge, preventing the identification of star overdensities, or because the two clusters are not actually companions {and, most likely, given the rotation of the Galaxy, they are destined to separate rapidly. We have to take into account that the two clusters are separated by a few arcminutes in projection, but are more than 1.5 kpc away in distance.  This possibility opens intriguing avenues for exploring their evolutionary pathways within the broader Galactic context.}\\

Furthermore, Garro~03's origin is not confirmed since we need to derive the radial velocities, which will allow us to reconstruct its orbits (e.g. \citealt{Garro2023}) and chemical abundances, which in turn will help us to understand if it is an in situ or an accreted object (e.g. \citealt{2024arXiv240508963M}). This ambiguity arises from its proximity to Garro~01. \cite{Garro_2020} suggested that Garro~01 may be associated with the Monoceros Ring (MRi, also known as the  Galactic anti-centre stellar structure;  \citealt{Newberg_2002}). The origins of the  MRi  may be a remnant of the tidally disrupted Canis Major dwarf galaxy \citep{2005MNRAS.362..906M,2005ApJ...626..128P,Morganson_2016,10.1093/mnras/stx3048} or it could be a result of the Galactic warp and flare, comprising stars from the MW disk displaced due to interactions between a satellite galaxy and the disk \citep{2006A&A...451..515M,Kalberla_2014,Sheffield_2018}.
However, \cite{2023MNRAS.526.1075P} excluded the association of Garro~01 with the MRi since MRi models predict larger distances $D>20$ kpc than have been inferred for the cluster, and the observed metallicity distribution of MRi is more metal-poor than Garro 01. In addition,   Garro 01's radial velocity ($\sim 143$ km s$^{-1}$) shows an offset with respect to what is expected from the MRi models ($\sim 0 - 100$ km s$^{-1}$). Therefore, given the similarities with Garro 01, it is plausible that Garro~03 could be a disk cluster that was created in situ, and has endured significant dynamical processes associated with the Galactic warp.

\section{Conclusions}
\label{sec:conclusions}
We report the discovery of another star cluster in the VVVX footprint, named Garro~03. It is located in the MW disk at a heliocentric distance of 14.1 kpc, and a galactocentric distance of 10.6 kpc below the Galactic plane at Z $= -0.89$ kpc. We found this cluster as a clear red giant overdensity in the same tile as another GC candidate that was recently discovered. On the basis of our photometric analysis, Garro~03 shows an age of $\sim$3 Gyr and [Fe/H]~$=-0.5\pm 0.2$ (Table \ref{table}). However, its nature is still under discussion;  it may be classified  as an old OC or a young GC.\\ %, therefore, it will be disentangled later on with spectroscopic follow-up of their members, which will help us to understand if it is an in situ or accreted object. \\
The recent discovery of this new cluster highlights the potential for uncovering numerous hidden star clusters within the MW’s poorly studied regions. These areas pose challenges for scrutiny due to the dust barrier and high stellar density. However, with upcoming instruments, for example  MOONS, 4MOST, the Vera Rubin Telescope, and JWST, significant strides forward are expected.

%%%%%%%%%%%%%%%%%%%%%%%%%%%%%%%%%%%%%%%%%%%%%%%%%
%%%%% ACKNOWLEDGEMENTS
%%%%%%%%%%%%%%%%%%%%%%%%%%%%%%%%%%%%%%%%%%%%%%%%%
\begin{acknowledgements}
%The authors are grateful to the anonymous referee for providing helpful comments and suggestions, which have improved the content of this paper.
We gratefully acknowledge the use of data from the ESO Public Survey program IDs 179.B-2002 and 198.B-2004 taken with the VISTA telescope and data products from the Cambridge Astronomical Survey Unit.  This work presents results from the European Space Agency (ESA) space mission \textit{Gaia}. \textit{Gaia} data are being processed by the {\it Gaia} Data Processing and Analysis Consortium (DPAC). Funding for the DPAC is provided by national institutions, in particular the institutions participating in the \textit{Gaia} MultiLateral Agreement (MLA). The \textit{Gaia} mission website is \url{https://www.cosmos.esa.int/gaia}. The \textit{Gaia} archive website is \url{https://archives.esac.esa.int/gaia}. \\

E.R.G. gratefully acknowledges the ESO Fellowship program. GF-T gratefully acknowledges the grants support provided by ANID Fondecyt Iniciaci on No. 11220340, ANID Fondecyt Postdoc No. 3230001 and CAS-ANID/CAS220009 (sponsoring researcher in both), from the Joint Committee ESO-Government of Chile under the agreement 2021 ORP 023/2021 and 2023 ORP 062/2023. R.K.S. acknowledges support from CNPq/Brazil through projects 308298/2022-5, 350104/2022-0 and 421034/2023-8. TP acknowledges support from the Argentinian institution SECYT (Universidad Nacional de Córdoba).
\end{acknowledgements}

%%%%%%%%%%%%%%%%%%%%%%%%%%%%%%%%%%%%%%%%%%%%%%%%%
%%%%% BIBLIOGRAPHY
%%%%%%%%%%%%%%%%%%%%%%%%%%%%%%%%%%%%%%%%%%%%%%%%%
\bibliographystyle{aa.bst}
\bibliography{references_thesis}

\begin{appendix}
\section{Statistical significance of peaks in 2D data using kernel density estimation}
\label{appendix:A}
{As explained in Sect. \ref{sec:Methodoly and results}, to identify statistically significant peaks in a two-dimensional dataset, we employed the KDE method to estimate the probability density function. The process involved several key steps: (i) data and grid setup: we used RA and Dec in degrees unit, and we create a grid over the range of these values to evaluate the KDE; (ii) we calculated the KDE to estimate the density function of the data. The choice of bandwidth in KDE is critical as it influences the smoothness of the density estimate. We tested bandwidths ranging from 0.05 to 0.3 to determine the optimal value; (iii) the local root-mean-square (RMS) calculation: to assess the local noise level around the peak, we calculated the RMS of the density values in a region surrounding the peak. Various sizes of the local region were tested to find the optimal configuration; finally, (iv) significance calculation: the significance of the peak was computed as the ratio of the peak value to the local RMS. This ratio indicates how prominently the peak stands out against the background noise.\\
The analysis yielded the following optimal parameters and their corresponding metrics: best bandwidth of 0.05, peak value of 85.32, local RMS of 83.17, significance of 1.92.
These results indicate that the optimal bandwidth of 0.05 provided a detailed yet smooth KDE, highlighting the peak effectively. The local RMS accurately captured the noise level near the peak. The significance value of 1.92 suggests that the peak is nearly twice the local noise level, indicating a statistically significant and prominent feature in the data.

The high significance value demonstrates the robustness of this methodology in distinguishing meaningful data features from background noise. }

\section{Table and figures}
\begin{table}[htpb]
%\onecolumn
\centering 
\renewcommand{\arraystretch}{1.1}
\caption{Garro~03 physical parameters.}
\begin{tabular}{lc}
\hline\hline
Parameters        &  Values\\
\hline
RA (J2000) &  14:01:29.3\\
DEC (J2000) & $-$65:30:57.0\\
$\mu_{\alpha}^{\ast}$ [mas yr$^{-1}$] & $-4.57 \pm 0.29$\\
$\mu_{\delta}$ [mas yr$^{-1}$] & $-1.36 \pm 0.27$\\
$V_{T}^{RA}$ [km s$^{-1}$]& $-305.7$\\
$V_{T}^{Dec}$ [km s$^{-1}$]& $-90.92$\\
$A_{G}$ & $1.54\pm 0.02$\\
$A_{Ks}$ & $0.25\pm 0.04$\\
D [kpc] & $14.1 \pm 0.5$ \\
$R_G$ [kpc]& $10.6$ \\
Z [kpc] & $-0.89$\\
$$[Fe/H]$$ & $-0.5\pm 0.2$\\
Age [Gyr]& $3\pm 1$ \\
$M_{Ks}$ & $-6.32\pm 1.10$\\
$M_V$ & $-4.06$ \\
$r_c$ [pc] & $3.07\pm 0.98$\\
$r_t$ [pc] & $19.36\pm 15.96$\\
c & 0.80\\
\hline\hline
\end{tabular}
\label{table}
\end{table}

\newpage

\begin{figure*}[htpb]
\centering
\includegraphics[width=18cm, height=6cm]{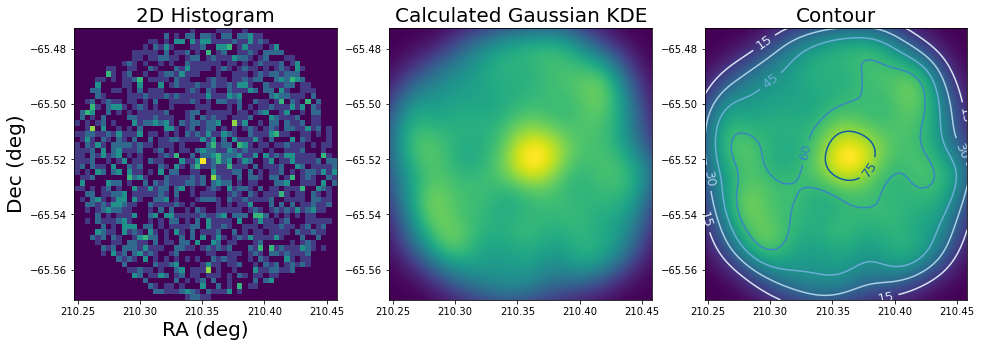} 
\includegraphics[width=18cm, height=6cm]{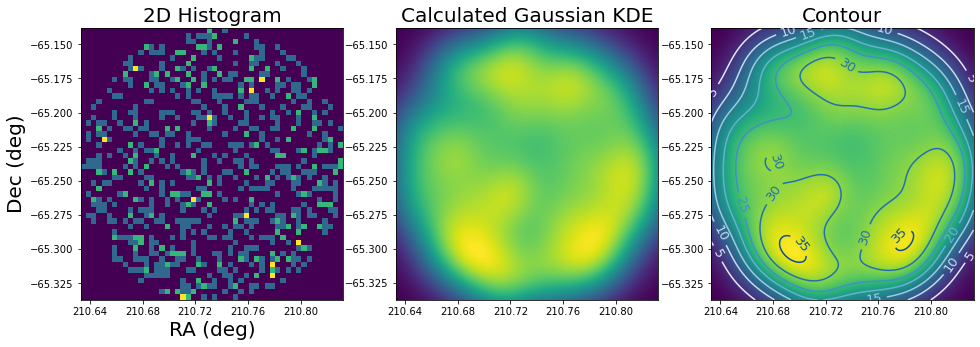} 
\caption{KDE spatial distribution. Top panels: Spatial distribution for Garro~03 selected at $r\lesssim 3'$ from the cluster centre, where we can clearly see a circular overdensity of stars at the cluster position. Bottom panels: Spatial distribution of a $3'$ field, selected away from the cluster field, for which no central overdensity is visible. For both cases, the samples used VVVX data, cleaned by nearby stars. These panels show the following from left to right: the 2D histogram and the calculated Gaussian KDE on  which we overplotted iso-density contours.}
\label{fig:KDEposition}
\end{figure*}

\begin{figure*}[htpb]
\centering
\includegraphics[width=15cm, height=5cm]{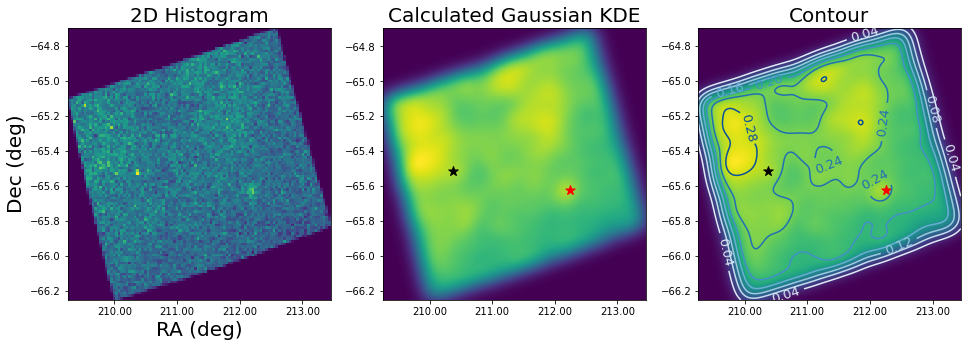} 
\caption{Spatial distribution of the entire tile $e0656$, where Garro~01 (red star symbol) and Garro~03 (black star symbol) are located. We apply the KDE technique as done in Fig. \ref{fig:KDEposition} with the main intention of detecting (if present) signatures indicating a bridge or streams between the two clusters. }
\label{fig:companion}
\end{figure*}

\end{appendix}

\end{document}